# Pricing the Information Quantity in Artworks

Lan Ju

Associate Professor in Finance

Peking University HSBC Business School

Zhiyong Tu

(Corresponding Author)

Associate Professor in Economics

Peking University HSBC Business School

zytu@phbs.pku.edu.cn

Changyong Xue

Ph.D. in Economics

Peking University HSBC Business School

**November. 2020**

# ABSTRACT

In the traditional art pricing models, the variables that capture the painting's content are often missing. Recent research starts to apply the computer graphic techniques to extract the information from the painting content. Most of the research concentrates on the reading of the color information from the painting images and analyzes how different color compositions can affect the sales prices of paintings. This paper takes a different approach, and tries to abstract away from the interpretation of the content information, while only focus on measuring the quantity of information contained. We extend the concept of Shannon entropy in information theory to the painting's scenario, and suggest using the variances of a painting's composing elements, i.e., line, color, value, shape/form and space, to measure the amount of information in the painting. These measures are calculated at the pixel level based on a picture's digital image. We include them into the traditional hedonic regression model to test their significance based on the auction samples from two famous artists (Picasso and Renoir). We find that all the variance measurements can significantly explain the sales price either at 1% or 5% level. The adjusted R square is also increased by more than ten percent. Our method greatly improves the traditional pricing models, and may also find applications in other areas such as art valuation and authentication.

## 1. Introduction

Artworks have become an increasingly important category in the global investors' portfolios. During the past twenty years, the annual total amount of artworks auctioned worldwide more than quadrupled while the world GDP only doubled.[1] For investors, how to price the artworks has always been the central question. The main difficulty stems from the fact that the artists place much emphasis on the uniqueness of their works and strongly avert the standardization. So the controversies concerning the interpretation and the valuation of artworks are just unavoidable.

The traditional methodologies on art pricing mainly fall into three big categories: hedonic, repeated sales and hybrid models.[2] In the hedonic model, the sales price of a painting is usually explained by the multiple factors ranging from painting attributes such as size, material and signature to sales conditions such as year, salesroom and sales location, etc (e.g.,

---

[1] *The Art Market in 2019*, AMMA and Artprice.com.

[2] Different from the hedonic model that attributes an implicit price to each of the time-invariant and time-varying characteristics of the item, the repeated sales model explicitly controls for differences in quality between artworks by considering only the items sold at least twice (Goetzmann, 1993; Pesando, 1993). The hybrid model combines the features of both the hedonic and the repeat sales models (Mei and Mose, 2002; Park et al., 2016).

Buelens and Ginsburgh, 1993, Chanel, 1995 and Taylor and Coleman, 2011, among others). Notice that the most important variable, the variables that capture the painting's content information, are often missing in the model. This has long been a drawback of the traditional models. After all, we buy paintings not for the frame but for what is on the canvas.

To control for the content heterogeneity, some literature introduces the content dummies into the hedonic model, such as Biey and Zanola (2005), Lazzaro (2006), Renneboog and Spaenjers (2013) and Ginsburgh et al. (2019). For example, Renneboog and Spaenjers (2013) categorized their sample into eleven groups by the distinct subject matters, and accounted for their influence on the sales price using Topic dummies.[3] However, these topic dummies only give a very limited description of the painting content. Even within the same category, the specific content of each painting, consequently, its sales price, can vary greatly. The following is a real example.

We notice that two Picasso's paintings, "*Homme À La Pipe Et Nu Couché, 1967*" and "*Nu Couché, 1967*", have the same physical features (length 146cm, width 114cm and oil canvas). They were sold in the same year of 2011 in the Sotheby's, London, but the former one obtained 4,801,250 GBP, two times that of the latter (2,281,250 GBP). Both paintings were created in 1967 hence belonging to the same period of Picasso. They also depict the same subject matter. The main difference is that the former contains more human figures than the latter.

This example shows that the topic dummies may be too coarse to effectively differentiate painting contents. If we can use better information variables than just dummies, we can price the painting more accurately. Nowadays we can easily access the digital images of paintings in the major databases. Then with the help of computer graphic techniques, we can withdraw much more information from these images.[4] So researchers start to introduce computer graphics into the art pricing models. Most recent research focuses on the reading of the color information from the painting images and tries to analyze how different color compositions can affect the sales prices of paintings.

For example, Pownall et al. (2016) find a price premium for paintings with higher color intensity based on the sample of Andy Warhol. Stepanova (2019) shows that the percentage of blue or orange color in a painting can significantly and positively affect the sales price with a Picasso sample. Ma et al. (2019) adopt a large sample mixed with different artists and reveal that the higher percentage of red or blue color in the painting can lead to a price premium.

---

[3] The eleven categories are: ABSTRACT, ANIMALS, LANDSCAPE, NUDE, PEOPLE, PORTRAIT, RELIGION, SELF-PORTRAIT, STILL_LIFE, UNTITLED, AND URBAN.

[4] Computer graphics is a discipline that studies the representation and manipulation of image data via computer algorithms.

The above research builds on the theory that the colors in a painting influence the human emotions which in turn affect the sales price. However, we argue that this color-emotion-purchase mechanism is not straight forward and still remains to be tested. Which colors relate to what emotions and how emotions link to the purchase behaviors are not clear. Especially, the empirical results seem not to be robust either. The colors that can produce a price premium differ due to different testing samples (artists) as shown above.

This paper tries to apply the computer graphics to the art pricing research from a very different perspective. Generally speaking, the previous research concentrates on reading the *meaning* of the painting information by computer algorithms. However, we hold that the connection between the computer interpretation of the painting and its final sales price can hardly be robust as the appreciation of artworks is a quite individualized experience. Therefore, we try to abstract away from the interpretation of the painting information, while only focus on measuring the *quantity* of the painting information.

To draw an analogy, we price a dish not by the consideration of its taste because people have individualized tastes. Instead, we price it by the amount of materials included because they are the common component for any dish. The more materials are contained in the dish, the higher the dish shall be priced. This approach may provide a more robust inference for the sales price than trying to probe into the subjective tastes. Note that in this analogy the amount of materials does not correspond to the actual costs of paint, frame or canvas for the painting because they are so marginal to the painting's sales value. They correspond to the amount of information in the painting as the information conveyed is exactly where the value of the painting hinges on.

In the information theory, Claude Elwood Shannon proposes using entropy to measure the amount of information a signal/event carries (Shannon, 1948). Entropy does not tell anything about what the signal/event information is about. Rather, it tells the amount of information the signal/event transmits. Shannon denominated the common unit of information as bit. Regardless of the meanings from different pieces of information, they can all be measured by bit for its information quantity. Similarly, we may also focus on the information quantity contained in a painting while completely abstain from the interpretation of its meaning.

Painting is a media for the artist to communicate with the audience. The more information is delivered, the more valuable the painting will be given other things equal. However, we do not know if any measurement for the information quantity contained in a painting has ever been suggested in literature. Under such a circumstance, we make a bold extension of Shannon entropy to the painting's scenario, and suggest using the variances of a painting's composing elements, i.e., line, color, value, shape/form and space, to measure the amount of information in the painting. These measures are calculated at the pixel level based on a picture's digital image.[5]

We introduce these measurements of information quantity of the painting into the hedonic model and test their significances using the auction sample of Pablo Picasso (1881-1973) from year 2000 to 2018. Picasso's artworks are a constant research topic in the art investment literature (e.g., Czujack et al., 2004, Forsund et al., 2006, Pesando and Shum, 2007 and Scorcu and Zanola, 2011). One reason is that Picasso is both prolific and versatile; his paintings in the market not only constitute a large sample, but also range across diverse styles. Another reason may be that Picasso's prints are actively traded and fairly liquid so the formation of their prices is highly marketized (Pesando and Shum, 1999 and Biey and Zanola, 2005).

Our main finding based on the sample of Picasso is that all our measurements of information quantity can significantly explain the sales price of the painting. These new information variables improve the adjusted R square of the hedonic model by more than ten percent. Our subsequent robustness test also carries out the same nalysis for the French impressionism artist Pierre-Auguste Renoir (1841-1919), and the result remains robust. All the information quantity variables are significant either at 1% or 5% level.

This paper suggests a new perspective to apply the computer graphics to the art pricing research. It does not ask the computer to simulate how human appreciates a painting. It requests the computer to do what it is good at, i.e., to draw the information quantity from a painting. Our measurements of information quantity are calculated by algorithm and free of personal judgment. They are easy to implement and can serve as robust new variables of information for all paintings. Not only do they improve on the hedonic model, but they can also serve as the new controls for the repeated sales and hybrid models. From a practical point of view, our measurements of information quantity of a painting may also assist in both valuation and authentication of paintings.

The reminder of the paper is organized as follows. Section 2 explains our theoretical framework. Section 3 describes the measurements of information quantity. Section 4 presents

---

[5] We will elaborate on these measurements in the later sections.

the data and the empirical methodology. Section 5 presents the results and Section 6 concludes.

**2. Theoretical Framework**

Fundamentally, the value of an artwork lies on the information it transmits. The more information a painting can carry, the bigger value it has hence shall be priced higher. But what "more information" means here seems ambiguous. For example, *Composition with Red Blue and Yellow* is a 1930 painting by Piet Mondrian. The whole picture consists of just a few straight lines and colored squares. In this painting, some audience may find fairly rich information while some may only see quite little in it. This is because people appreciate the painting differently.

Appreciation of a painting is a subjective experience. Computational esthetics is a strand of research that tries to *understand* how people appreciate the information in a painting with computer algorithms. As we showed before, the connection between the computer's understanding and the painting's price is not robust. The color-emotion-purchase mechanism in the previous literature is one of such theories. But this mechanism is neither fully understood nor rigorously tested yet.

Therefore, our research chooses to abstract away from the understanding of the painting. Instead, we only consider the amount of data being contained in the painting's content. We call this kind of data amount in a painting as the *information quantity*. That is also how we define "more information" in this paper. More information in a painting involves no personal judgment; it just means a bigger amount of data, i.e, a larger information quantity.

Then how do we quantify the information quantity that is independent of the painting interpretation? The idea is derived from the concept of Shannon entropy. Shannon held that the information is used to resolve uncertainty. Then the amount of information a signal/event transmits can be quantified by the degree of uncertainty the signal/event entails. The higher uncertainty from a signal/event implies a bigger chance of "surprise" by the receiver hence conveying more information. More rigorously, Shannon entropy is defined as the sum of the probability of each state times the log probability of that same state, i.e., $-\sum_{i=1}^{n} p_i log p_i$, where $p_i$ is the probability of state *i* among all the *n* states.

However, there is no direct counterpart of Shannon entropy for the measurement of the information quantity in the painting. Because entropy is defined over the probability distributions, while a painting is a realized sample. To explain this, we can think of a painting as a building block made by thousands of pixels. The artist decides on each pixel for its attributes such as line, color and value etc. If the artist's decision is a random choice, then each possible choice may be assigned a probability. The successive decisions on all pixels finally produce the whole picture. In this sense, every pixel is an uncertain event, and the whole picture is a joint event of all pixels. So the entropy of the painting must be positively related to the total number of pixels, which is just the picture's resolution.

Obviously, such entropy is not what we are looking for because it measures the amount of information an unpainted image could *possibly* deliver given its resolution. But what we need here is the measurement of information quantity for an existing painting, i.e., a realized sample of pixels. Nevertheless, the concept of entropy points to the uncertainty as the direction for exploration of the potential measure. For a realized sample, the inherent uncertainty produces the actual volatility. So we can reason that the counterpart of entropy for a realized sample must be the sample variance.

In the painting's scenario, we could use the variances of a painting's composing elements to measure the amount of information delivered. These variances are calculated at the pixel level from the digital image of each painting. The larger these variances are, the more information the painting carries. For example, a larger variance of line in a painting provides a more complex image implying richer information. Considering that the variance represents the deviation from the mean, we can say that the larger such deviation may reflect more uncertainties/volatilities, consequently bringing more information.

A painting is composed of seven common elements, i.e., line, color, value, shape/form, space and texture.[6] Lines are marks in the painting, and shape/form is a two/three-dimensional design encased by lines. Color is an element consisting of hues, while value is the perceivable lightness or darkness of hue. Space refers to the perspective distance between and around shapes and objects. Finally, texture describes the surface quality of the work. We will use material dummies to control for the texture in our subsequent analysis. So we leave it out of our variance measurement here. Finally, we can propose that the information quantity of a painting is defined by the following log linear equation (1), where $V(\cdot)$ represents the variance function of each element:

$$\text{Information Quantity} = V(\text{line}) \times V(\text{color}) \times V(\text{value}) \times V(\text{shape/form}) \times V(\text{space}) \qquad (1)$$

The above specification could also be justified from the perspective that the information is the resolution of uncertainty. By this logic, we could think of the information quantity as a joint probability distribution (aggregation) over all its arguments, then it shall take the log linear form.

---

[6] Please see https://en.wikipedia.org/wiki/Elements_of_art.

**3. Measurements of Information Quantity**

Next, we explain how we will measure each variance in Equation (1). Actually, all our variance measurements are carried out at the pixel level of each picture. Note that the basic unit of a digital image is called pixel. More pixels in a given area imply a higher resolution of the image. The resolution is determined by the numbers of pixels in both horizontal and vertical directions. In Figure 1, if we enlarge the crossing point of two lines in a picture (Picture A), we can see that they are actually made up of series of black and white pixels (Picture B).

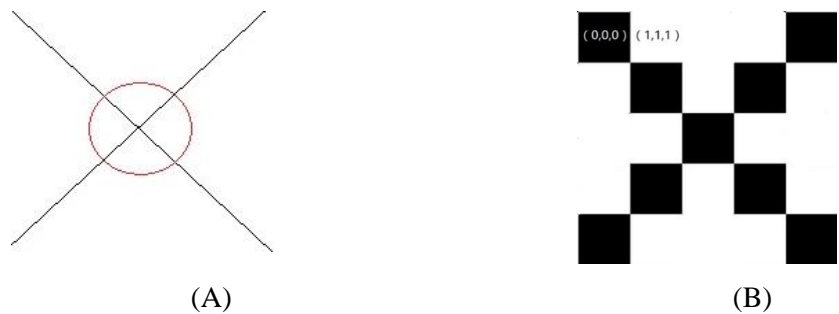

(A)                  (B)

**Figure 1. Representation of Pixels in a Digital Image**

Each pixel can be characterized by its unique hue and location in the image. The arrays of pixels constitute lines, and the combinations of hues form colors. In Section 3.1, we will describe our variance measurements in a heuristic manner.

3.1 The Measurements

The variances of a painting's composing elements are measured at the micro level using its digital image, i.e., in the form of pixels. At such level, the lines and colors become dots covering each segment of pixel, and then we can evaluate the variations of these segmental dots.

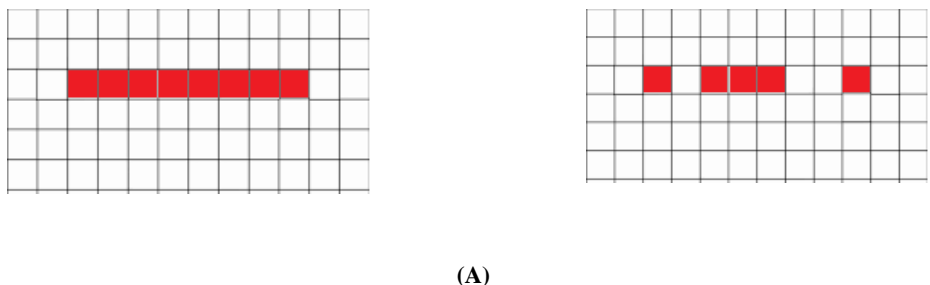

**(A)**

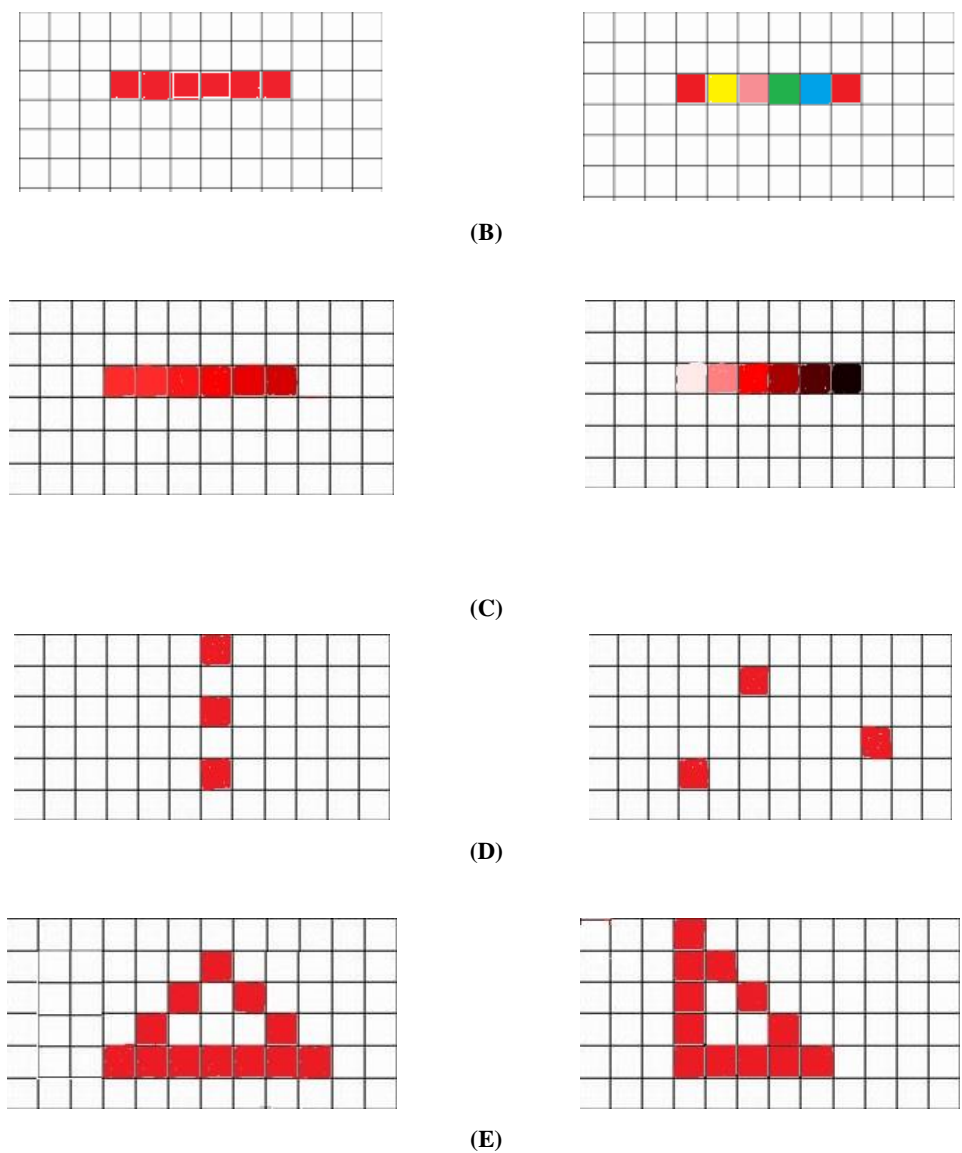

**(B)**

**(C)**

**(D)**

**(E)**

**Figure 2. Representation of the Variations of Line, Color, Value, Shape and Space at the Pixel Level**

In Figure 2 (A), a line crosses seven consecutive pixels in the left picture; while in the right picture it takes a break after crossing the first pixel, and takes two breaks after crossing the fifth. Intuitively, the variation of lines in the right picture is larger. So it provides more information according to our definition. As all the painting images under analysis are the color ones, we first need to graylize the picture and find the resulting image's line contour, and then calculate the contour's variance.

Figure 2 (B) shows two lines that consist of pixels with different colors. The pixels of the left line are consistently red, while those of the right are of varying colors. So the color variation of the left picture is larger. For a whole picture, our calculation involves all the colors of all pixels according to a color variance formula.

In both pictures of Figure 2 (C), the red pixels increase their degrees of darkness from the left to the right along the line. It is easy to see that the line in the right picture has more variations of lightness/darkness (called value) than the left one. In fact, every pixel in a painting has a defined value number; therefore we can calculate the variance of value for the whole painting.

In the left picture of Figure 2 (D), the dots are even distributed along a vertical line. While in the right picture, these dots are more scattered, and the distances among them vary a lot. So the right picture has a larger variation in term of space. The variance of space can be calculated using the x-y coordinates of each pixel positioned in the whole painting.

Finally, we look at the variance of shape/form. Figure 2 (E) depicts two triangles made up of pixels. If we view them laterally and consider their shapes from an integral perspective, we could say that the left triangle has a smaller variation in term of its shape because it is more laterally symmetric than the right one.[7] As two and three dimensional designs have no distinction at the pixel level, we will just use shape rather than shape/form hereinafter.

3.2 The Measurement Formulae

In this section, we will introduce the detailed formula for each variance measurement. They mostly follow the literature in computer graphics (e.g., Sural et al., 2002; Coleman et al., 2005; Alain H. and Ziou, D., 2010; Wang, S., 2012). The calculation relies on the digital color image of each painting drawn from the database. Therefore, we first need to know how to characterize color in a picture.

The hue of a pixel can be described by a RGB system, where RGB stands for the proportions of three primary colors, red, green and blue, in a pixel respectively. For example, RGB (0,0,0) in Figure 1 (B) means the proportions of red, green and blue are all 0%, so the pixel is black. While, RGB (1,1,1) indicates the proportions of three colors are 100%, then the pixel shows as white.

*I. The variance of line*

In order to obtain the variance of line of an image, we need to follow three steps as illustrated in Figure 3.

---

[7] We only consider the lateral symmetry in this paper due to our visual habit when viewing a painting.

First, we convert a colorful picture into a gray one via the floating-point algorithm.[8] In Figure 3, we can see that the black and white Picture B is produced by this algorithm from the original colorful Picture A.

Second, we apply the edge detection method to obtain the image's line structure attribute.[9] Picture C in Figure 3 is generated from Picture B by the edge detection algorithm.

Finally, the variance of line is calculated based on the grayscales of those edges after the edge detection treatment according to the following formula:

$$V(line) = \frac{\sum_i^N (grayscale_i - \mu)^2}{N} \qquad (2)$$

Note that the grayscale of each pixel *i* is reduced to {0,1} by the edge detection procedure. $\mu$ is the average of all the grayscales of pixels and *N* is the total number of pixels in the edge detection image of a painting.

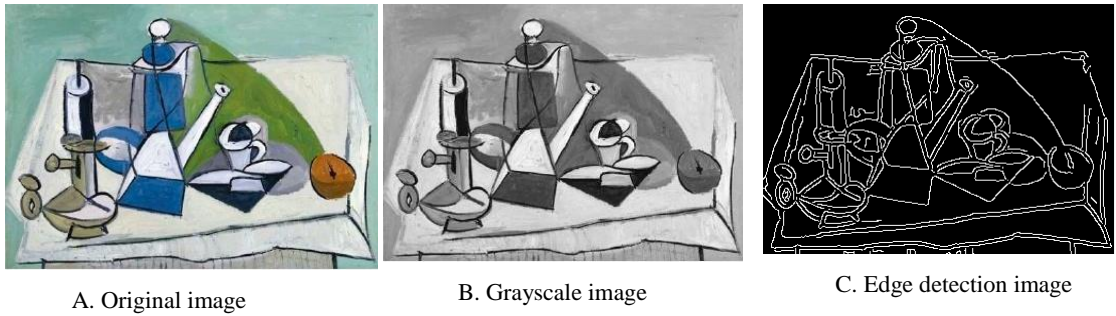

A. Original image　　　　B. Grayscale image　　　　C. Edge detection image

**Figure 3 . Three Steps to Calculate the Variance of Line**

*II. The variance of color*

The variance of color is defined as the variance of the hues of all pixels in the image of a painting. As we describe above, the hue of a pixel can be characterized by a three-dimensional RGB system. While, the hue value is a one-dimensional degree number that is reduced from the RGB to specifically represent the unique color of a given pixel.

---

[8] The floating-point algorithm calculates a grayscale for a pixel with certain RGB by the formula: grayscale=R*0.3+G*0.59+B*0.11. The grayscale determines a relative distance (or grayness) between white and black, and the color of the pixel is transformed into gray based on this scale.

[9] Edge detection is a basic method in image processing and computer vision, whose purpose is to identify the points with obvious grayscales changes in an image. Edge detection reduces the amount of data by removing the irrelevant information from the image, while retaining the image's important structural attributes.

The hue value is defined to range from 0° to 360°, starting from the red and go in an anti-clockwise direction. Red is 0°, green is 120°, and blue is 240° and so on.[10] Then the variance of hue values can be defined as:

$$V(color) = \frac{\sum_i^N (huevalue_i - \mu)^2}{N} \tag{3}$$

where $huevalue_i$ is the hue value of pixel $i$ and $\mu$ is the average of all the hue values of pixels and $N$ is the total number of pixels in the painting image.

*III. The variance of value*

The value of pixel $i$ measures the lightness or darkness of the pixel's hue. So the value of pixel $i$ can also defined by its RGB of the corresponding hue according to the following formula:

$$value_i = \sqrt[2.2]{\frac{(\frac{R}{255})^{2.2} + (1.5\frac{G}{255})^{2.2} + (0.6\frac{B}{255})^{2.2}}{1 + 1.5^{2.2} + 0.6^{2.2}}} \tag{4}$$

Then the variance of value for the whole painting can be defined as:

$$V(value) = \frac{\sum_i^N (value_i - \mu)^2}{N} \tag{5}$$

where $value_i$ is the value of pixel $i$ from formula (5), $\mu$ is the average of all pixels' values and $N$ is the total number of pixels in the painting image.

*IV. The variance of space*

As we mention earlier, the digital image of a painting is a X x Y matrix of pixels, where X and Y are the total numbers of pixels from the horizontal and vertical directions respectively. So we can number each pixel horizontally from 1 to X and vertically from 1 to Y. Then the spatial position of pixel $i$ is uniquely identified by its coordinates ($x_i$, $y_i$), where $x_i$ and $y_i$ belong to {1,2,…X} and {1,2…Y} respectively. We can calculate the variances of x and y coordinates respectively, and then use their weighted sum to define the variance of space.

---

[10] Let R,G,B be the three proportion numbers in the RGB system, and max and min be the maximum and minimum of R, G, and B, then the hue value of a pixel can be defined as: $60° \times \frac{G-B}{max-min} + 0°$ if if max = R and G ≥ B ; $60° \times \frac{G-B}{max-min} + 360°$, if max = R and G < B ; $60° \times \frac{B-R}{max-min} + 120°$, if max = G ; $60° \times \frac{B-R}{max-min} + 120°$, if max = G; and undefined if max=min.

Specifically, we first normalize each coordinate $\hat{x}_i = \frac{x_i}{X}$ and $\hat{y}_i = \frac{y_i}{Y}$ so as to reduce their values to the range of zero and one. Then graylize the image and carry out the edge detection. After that, the remaining pixels become only black and white, and we calculate the variance of space for those remaining black pixels. The spatial variance as the weighted sum of the variances of x and y coordinates is defined by equation (6):

$$V(space) = \frac{\sum_i^N (\hat{x}_i - \mu_x)^2}{2X} + \frac{\sum_i^N (\hat{y}_i - \mu_y)^2}{2Y} \qquad (6)$$

where $\mu_x$ and $\mu_y$ are the average values of all $\hat{x}_i$ and $\hat{y}_i$ respectively.

*V. The variance of shape*

The variance of shape involves more clarification. Initially, we wish to detect the total number of different shapes that appear in a picture at the pixel level. But we find no reference in literature for such a task. In fact, we think it may be impossible to accomplish this task with computer algorithms because defining what a shape is involves much subjective judgment. In a sense, anything can be seen as a shape. So we choose to consider the whole picture as one big complex shape. Then its shape variance can be measured by the inverse of the degree of lateral symmetry of the image. Note that we only consider the lateral symmetry of the painting here because we think it conforms to our visual habit.

We first need to partition each picture into two equal halves: the left half is denoted as L and the right half as R. Then calculate the structural similarity of L and R. For this, we also need to graylize the image, and find the grayscale for each pixel. The degree of lateral symmetry is defined as:

$$SSIM_{(L,R)} = \frac{(2\mu_L \mu_R + c_1)(2\sigma_{LR} + c_2)}{(\mu_L^2 + \mu_R^2 + c_1)(\sigma_L^2 + \sigma_R^2 + c_2)} \qquad (7)$$

where $\mu_L$ and $\sigma_L$ are the mean and variance of grayscales of all pixels in the left half of the picture; and $\mu_R$ and $\sigma_R$ are the right half counterparts. $\sigma_{LR}$ is the covariance of the grayscales of pixels in L and R. $c_1 = (k_1 L)^2$ and $c_2 = (k_2 L)^2$ are all constants used to maintain stability, where $k_1 = 0.01, k_2 = 0.03, L = 256$.

Then we define the variance of shape as the scaled inverse of SSIM, where the scaling aims to restrict the measurement to the range of zero and one:

$$V(shape) = \frac{1}{1000 * SSIM_{(L,R)}} \qquad (8)$$

3.3 An illustrative Example

The above description of these mathematical formulae seems pretty abstract. So in the following Figure 4, we use two pictures of the same resolution but different content to illustrate how our measurements of variances can reflect their differences in the amount of information conveyed. The two roosters in Figure 4 have visually distinct delicacies, and the calculated five measurements for the information quantity are presented in Table 1.

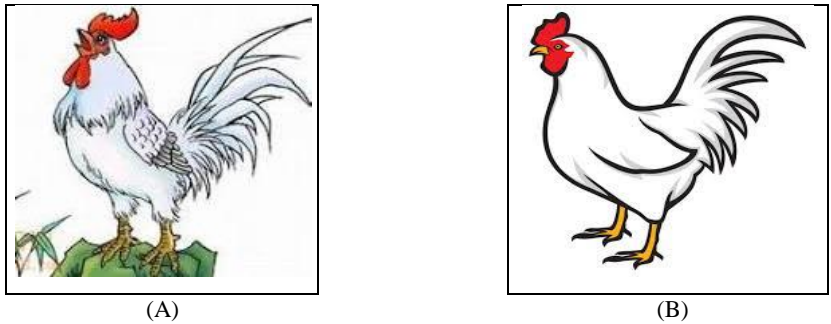

(A)                                    (B)

**Figure 4. Two Roosters with Different Delicacies and Information Quantities**

Based on our calculation, the variances of line, color and shape for the left rooster (A) are 68%, 41% and 50% higher than those of the right rooster (B) respectively. While for the other two variances, the differences are not so large. The variance of value for rooster (B) is 23% higher than that of rooster (A), while the variance of space for rooster (B) is only 3% higher than that of rooster (A). So intuitively, we may say that the composition of the left picture produces more variations overall. Consequently it contains more information.

To have an idea as to how these five dimensions of variances may impose an aggregate impact on the sales price, let's suppose these two roosters are drawn by Picasso with everything else equal except the painting content (i.e., the same subject, size, salesyear and salesroom, etc). Then we can plug their variance data into our estimated pricing function (on the Picasso sample) in the following Table 4.[11] The projection shows that the left rooster shall be priced 83% higher than the right one just because of its richer content information.

**Table 1. Five Measurements of Variances for the Two Roosters in Figure 4**

|            | V(line) | V(color) | V(value) | V(shape) | V(space) |
|------------|---------|----------|----------|----------|----------|
| Rooster (A)| 0.0841  | 0.297    | 0.207    | 0.003    | 0.062    |
| Rooster (B)| 0.0499  | 0.210    | 0.255    | 0.002    | 0.064    |

---

[11] Note that we need first align the resolutions of the pictures of two roosters to those in the Picasso sample.

## 4. Data and Empirical Methodology

We now describe our data source, empirical methodology, as well as the hypothesis for a more formal analysis.

4.1 Data Description

Our data are drawn from the Blouin Art Sales Index (BASI), presently the largest known database of artworks online that provides data on artworks sold at auction at over 350 auction houses worldwide (Korteweg et al., 2015). We adopt all the records of Picasso's paintings from year 2000 to 2018 except those without pictures or prices.[12] Each sales record includes such information as the title of the painting, digital image, size, date of creation, materials, date and city of sales, salesrooms, signature and prices, etc.[13] Altogether, we obtain 720 records. For the robustness analysis, we draw on the auction records of the same period from another famous French artist, Pierre-Auguste Renoir, whose data are also from BASI. As these digital images are all from the same database, their resolutions are controlled in the first place.

The following Table 2 provides the summary statistics of the five information quantity variables according to our definitions for the Picasso sample. The variables of V_line, V_color, V_value, V_shape and V_space represent the variances of line, color, value, shape and space respectively. All variables have the range of zero to one. From the following correlation matrix, we can see that these five dimensions of measures for the information quantity are weakly correlated in general.

**Table 2. Summary Statistics of the Information Quantity Variables**

| Variable | Obs | Mean | Std. Dev. | Min | Max |
|---|---|---|---|---|---|
| V_line | 720 | .093 | .019 | .04 | .156 |
| V_color | 720 | .213 | .086 | .006 | .441 |
| V_value | 720 | .225 | .059 | .035 | .398 |
| V_shape | 720 | .009 | .008 | .002 | .116 |
| V_space | 720 | .062 | .009 | .042 | .094 |

| Corr | V_line | V_color | V_value | V_shape | V_space |
|---|---|---|---|---|---|
| V_line | 1.000 | | | | |
| V_color | -0.094 | 1.000 | | | |
| V_value | -0.203 | 0.298 | 1.000 | | |
| V_shape | 0.681 | -0.147 | -0.239 | 1.000 | |
| V_space | 0.332 | 0.028 | -0.025 | 0.300 | 1.000 |

Table 3 summarizes all other variables not related to the painting content. Price is the hammer price of the painting sold in the auction denominated in USD. The most expensive painting in our sample is "*Les femmes d'Alger (Version 'O')*", sold for 179,365,000 USD in 2015 in New York; while the cheapest painting in the sample is "*Vase tapestry*", sold only for 1,280 USD in 2012 in Chicago. The mean price is 5,893,419 USD.

---

[12] Note that the records before 1997 in BASI do not provide the digital pictures, and the auction records between 1997 and 1999 are very sparse.

[13] Note that all the hammer prices are converted into U.S. dollars at the spot exchange rates at the time of sales.

Age is the number of years between a painting's time of creation and its time of sale. Salesyear is the year when the painting was sold. Surface is the area of the painting, and the two dummy variables, Signature and Dated, represent if the painting is signed or dated by the artist. Other control dummies include Material (painting material), City (auction location) and Salesroom (auction house). The summary statistics in Table 3 show that most of the Picasso paintings in our sample are on canvas, sold by either Christie's or Sotheby's in London and New York.

**Table 3. Summary Statistics of other Variables**

| *Variable* | *Obs* | *Mean* | *Sd* | *Min* | *Max* |
|---|---|---|---|---|---|
| Price($) | 720 | 5,893,419 | 1.20e+07 | 1,280 | 179,365,000 |
| Age | 720 | 74.68 | 20.63 | 47 | 125 |
| Salesyear | 720 | 2,010 | 5.203 | 2,000 | 2,018 |
| Surface(1000cm$^2$) | 720 | 6.378 | 10.67 | 0.0160 | 163.5 |
| Signature | 720 | 0.564 | 0.496 | 0 | 1 |
| Dated | 720 | 0.617 | 0.487 | 0 | 1 |
| **Material** | | | | | |
| Board | 720 | 0.047 | 0.212 | 0 | 1 |
| Burlap | 720 | 0.061 | 0.240 | 0 | 1 |
| Canvas | 720 | 0.815 | 0.388 | 0 | 1 |
| Cardboard | 720 | 0.022 | 0.147 | 0 | 1 |
| Ceramic | 720 | 0.033 | 0.180 | 0 | 1 |
| Others | 720 | 0.021 | 0.143 | 0 | 1 |
| **City** | | | | | |
| London | 720 | 0.358 | 0.480 | 0 | 1 |
| New York | 720 | 0.549 | 0.498 | 0 | 1 |
| Paris | 720 | 0.063 | 0.242 | 0 | 1 |
| Others | 720 | 0.031 | 0.172 | 0 | 1 |
| **Salesroom** | | | | | |
| Christie's | 720 | 0.518 | 0.500 | 0 | 1 |
| Sotheby's | 720 | 0.428 | 0.495 | 0 | 1 |
| Others | 720 | 0.054 | 0.226 | 0 | 1 |

4.2 Hedonic Model

The following is our benchmark regression of the hedonic model:

$$\log(p_{it}) = \sum_{j=1}^{5} \theta_j \log(A_{ji}) + \sum_{k=1}^{m} \alpha_k x_{ki} + \sum_{t=t_0}^{T} \varphi_t D_{it} + \varepsilon_i \tag{9}$$

$p_{it}$ is the price of painting *i* sold at time *t*. $A_{ji}$ represents the five information quantity variables V_line, V_color, V_value, V_shape and V_space for painting *i*, and they are specified as log linear as in Equation (1).[14]

Except for the variables of information quantity, the regression also includes other usual explanatory variables listed in Table 3. $x_{ki}$ is defined as the set of time-invariant characteristics of painting $i$ (e.g., the size, material and signature). $D_{it}$ is the set of time varying idiosyncratic attributes (e.g., year dummy) for painting $i$.

---

[14] As the five measurements of variances belong to (0, 1), we multiply their values by 1000 before the logarization.

4.3 Hypothesis

Based on the above argument, larger variances of line, color, value, shape and space imply more information delivered in the painting; hence its sales price shall be higher given other things equal. Then we expect that V_line, V_color, V_value, V_shape and V_space are all positively related to the sales price given other controls.

## 5. Regression Results

In this section, we carry out the benchmark regression and the robustness analysis.

5.1 Benchmark regression

With the Picasso sample, we run the benchmark cross-sectional regressions with robust standard errors for different model specifications. The results are provided in the following Table 4.

Table 4. Results of Benchmark Regressions

| VARIABLES | (1) log(p) | (2) log(p) | (3) log(p) |
|---|---|---|---|
| *Painting Information* | | | |
| log(V_line) | | 23.89*** | 24.23*** |
| | | (6.643) | (6.647) |
| log$^2$(V_line) | | -2.607*** | -2.728*** |
| | | (0.735) | (0.738) |
| log(V_color) | | 0.383*** | 0.306*** |
| | | (0.0783) | (0.0827) |
| log(V_value) | | | 0.459*** |
| | | | (0.172) |
| log(V_shape) | | | 0.242** |
| | | | (0.120) |
| log(V_space) | | | 0.925*** |
| | | | (0.329) |
| *Painting Attribute* | | | |
| Surface | 0.108*** | 0.104*** | 0.107*** |
| | (0.0119) | (0.0118) | (0.0120) |
| Surface$^2$ | -0.000642*** | -0.000617*** | -0.000643*** |
| | (9.33e-05) | (9.23e-05) | (9.54e-05) |
| Age | 0.00964*** | 0.0122*** | 0.0140*** |
| | (0.00280) | (0.00278) | (0.00280) |
| Signature | 0.0400 | 0.0316 | 0.0227 |
| | (0.101) | (0.102) | (0.102) |
| Dated | 0.335** | 0.319** | 0.329** |
| | (0.150) | (0.146) | (0.144) |
| Material | control | control | control |
| *Other Control* | | | |
| City | control | control | control |
| Salesroom | control | control | control |
| Salesyear | control | control | control |
| Constant | 8.643*** | -47.77*** | -53.36*** |
| | (0.719) | (15.14) | (14.92) |
| Observations | 720 | 720 | 720 |
| Adj-R-squared | 0.459 | 0.495 | 0.510 |

Robust standard errors in parentheses. *** p<0.01, ** p<0.05, * p<0.1

Column (1) is the result from the traditional hedonic model; column (2) introduces the variances of line and color into the model, and column (3) includes all five variance measures. Our results show that the five measurements of information quantity of the painting are all significant to explain the sales price at either 1% or 5% level. It confirms our hypotheses that the information quantity measured on the pixels can supply significant information for the painting's value. The total adjusted R square is also improved by more than ten percent after adding these information variables.

We observe that the estimated coefficient of the line variance is the biggest among all the five measures. It implies that the line, i.e., the contour of the picture, plays the most important role in transmitting the painting information. The quadratic term of V_line is significantly negatively related to the price, which indicates a decreasing marginal effect of line variance to the painting information hence the sales price.

The variances of color, value, shape and space are all significantly positively related to the sales price because the bigger variances of these variables indicate more information the painting delivers. Other control variables produce the usual results similar as those in the art pricing literature. For the brevity reason, we omit the regression outputs for some regular control variables throughout this paper. Please see the Appendix for the detailed regression outputs that include all controls.

*5.2 Robustness*

Next, we wish to see if our suggested measurements of information quantity also work for the paintings of other artists with different styles. So we further analyze the sales records for the French artist Pierre-Auguste Renoir, who is the major representative for the impressionism art school. The number of his auction records is also large enough for a robust inference.

The impressionism portrays overall visual effects instead of details, and constructs the picture from freely brushed colors that take precedence over line and contours. It is interesting to see whether the five measures we propose can also capture the information quantity for such a different type of paintings.

We obtain 1,147 records for Pierre-Auguste Renoir from BASI (see the Appendix for the summary statistics). The following Table 5 provides the regression outputs, where column (4) is the output from the combined sample of both Picasso and Renoir.

Table 5. Results for Pierre-Auguste Renoir and Two Artists Combined

| VARIABLES | (1) log(p) | (2) log(p) | (3) log(p) | (4) log(p) |
|---|---|---|---|---|
| *Painting Information* | | | | |
| log(V_line) | | 8.535* | 11.23** | 19.76*** |
| | | (5.159) | (5.460) | (4.423) |
| log(V_line$^2$) | | -0.888 | -1.237** | -2.270*** |
| | | (0.546) | (0.582) | (0.485) |
| log(V_color) | | 0.156*** | 0.147*** | 0.401*** |
| | | (0.0356) | (0.0365) | (0.0426) |
| log(V_value) | | | 0.429*** | 0.409*** |
| | | | (0.116) | (0.120) |
| log(V_shape) | | | 0.107** | 0.152*** |
| | | | (0.0495) | (0.0581) |
| log(V_space) | | | 0.837*** | 1.011*** |
| | | | (0.260) | (0.253) |
| *Painting Attribute* | | | | |
| Surface | 0.857*** | 0.836*** | 0.836*** | 0.148*** |
| | (0.0356) | (0.0356) | (0.0350) | (0.0133) |
| Surface$^2$ | -0.0354*** | -0.0345*** | -0.0344*** | -0.000914*** |
| | (0.00280) | (0.00268) | (0.00281) | (0.000121) |
| Age | 0.00395 | 0.00356 | 0.00246 | -0.000634 |
| | (0.00254) | (0.00248) | (0.00243) | (0.00155) |
| Signature | 0.481*** | 0.453*** | 0.471*** | 0.354*** |
| | (0.0722) | (0.0699) | (0.0696) | (0.0691) |
| Dated | -0.0503 | 0.0260 | 0.0242 | 0.841*** |
| | (0.173) | (0.168) | (0.166) | (0.0956) |
| Material | control | control | control | control |
| *Other Control* | | | | |
| City | Control | control | control | control |
| Salesroom | Control | control | control | control |
| Salesyear | Control | control | control | control |
| Constant | 8.931*** | -12.16 | -17.15 | -41.63*** |
| | (0.471) | (12.15) | (12.81) | (10.06) |
| | | | | |
| Observations | 1,147 | 1,147 | 1,147 | 1,867 |
| Adj-R-squared | 0.602 | 0.610 | 0.621 | 0.581 |

Robust standard errors in parentheses. *** p<0.01, ** p<0.05, * p<0.1

We can see that the variances of line, color, value, shape and space are all significant at 1% or 5% level no matter from the single or the combined sample. Also, the signs of their estimated coefficients are all the same as those from the Picasso sample. It shows that our measurements of information quantity are fairly robust across different artistic styles.

One obvious difference here is that the estimated coefficient of the line variance from the Renoir sample is much smaller compared to that of the Picasso sample. This result is easy to interpret. For the impressionism paintings that emphasize personal feelings, the marginal contribution of line or contour to the information quantity in the painting must be smaller than those paintings with more realistic styles.

# 6. Conclusion

The above analysis shows that introducing the measurements for the information quantity of the painting can be valuable to the art pricing research. These measurements include the variances of line, color, value, shape/form and space that are the basic composing elements of any painting. Our measurements do not try to interpret the meaning of the information that a painting transmits. Instead, they focus on the quantification of the amount of information the painting contains.

This paper applies the computer graphic techniques to the digital images of the auctioned paintings from two famous artists (Picasso and Renoir), and calculates the five measurements of information quantity for these paintings. Then we include them into the traditional hedonic regression model, and find all of they can significantly explain the sales price either at 1% or 5% level. The adjusted R square is also increased by more than ten percent.

To our knowledge, this is the first research that proposes to use the information quantity of the painting as a more underlying and universal predictor for the painting's sales price. It is at this key point that the paper differentiates itself from the computational aesthetics literature, which focuses on the application of computer algorithm to the interpretation of the painting information. Our empirical results show that this approach indeed can greatly improve the traditional art pricing models. The methodology may also find applications in other areas such as art valuation and authentication.

As the first attempt to extend the concept of Shannon entropy in information theory to the measurement of the information quantity of the painting, our suggested measures may need further refinements in the future studies. For example, how to find more elaborate measures for the shape and space variances may need further exploration. The current robustness analysis is also limited by our data availability, which may be strengthened in the future.

Nevertheless, we think introducing the concept of information quantity into the art pricing research is a fruitful direction. We can further extend our approaches to artworks of higher dimensions, e.g., the three-dimension. Then we can also study the pricing of other types of artworks, such as sculptures and porcelains.

# Appendix

*Tables with more detailed outputs*

**Table A1. The Regression Results with all Controls of Table 4**

| VARIABLES | (1) lprice | (2) lprice | (3) lprice |
|---|---|---|---|
| **Painting Information** | | | |
| log(V_line) | | 23.89*** | 24.23*** |
| | | (6.643) | (6.647) |
| log(V_line$^2$) | | -2.607*** | -2.728*** |
| | | (0.735) | (0.738) |
| log(V_color) | | 0.383*** | 0.306*** |
| | | (0.0783) | (0.0827) |
| log(V_value) | | | 0.459*** |
| | | | (0.172) |
| log(V_shape) | | | 0.242** |
| | | | (0.120) |
| log(V_space) | | | 0.925*** |
| | | | (0.329) |
| **Painting Attribute** | | | |
| Surface | 0.108*** | 0.104*** | 0.107*** |
| | (0.0119) | (0.0118) | (0.0120) |
| Surface$^2$ | -0.000642*** | -0.000617*** | -0.000643*** |
| | (9.33e-05) | (9.23e-05) | (9.54e-05) |
| Age | 0.00964*** | 0.0122*** | 0.0140*** |
| | (0.00280) | (0.00278) | (0.00280) |
| Signature | 0.0400 | 0.0316 | 0.0227 |
| | (0.101) | (0.102) | (0.102) |
| Dated | 0.335** | 0.319** | 0.329** |
| | (0.150) | (0.146) | (0.144) |
| Material | | | |
|   board | 1.912*** | 1.700*** | 1.741*** |
| | (0.574) | (0.530) | (0.531) |
|   canvas | 2.120*** | 1.914*** | 1.956*** |
| | (0.555) | (0.515) | (0.519) |
|   cardborad | 2.157*** | 1.944*** | 1.942*** |
| | (0.538) | (0.499) | (0.503) |
|   panel | 1.825*** | 1.622*** | 1.462*** |
| | (0.561) | (0.511) | (0.527) |
|   paper | 1.112** | 1.113** | 1.004* |
| | (0.547) | (0.516) | (0.521) |
| **Other Control** | | | |
| City | | | |
|   New York | 0.750** | 0.715** | 0.703** |
| | (0.369) | (0.353) | (0.350) |
|   London | 0.913** | 0.903** | 0.894** |
| | (0.374) | (0.355) | (0.350) |
|   Paris | 0.0490 | 0.228 | 0.164 |
| | (0.392) | (0.392) | (0.390) |
| Salesroom | | | |
|   Christie's | 0.662*** | 0.677*** | 0.622*** |
| | (0.207) | (0.207) | (0.207) |
|   Sotheby's | 0.601*** | 0.653*** | 0.606*** |
| | (0.207) | (0.207) | (0.208) |
| Saleyear | | | |
| 2001.year | -0.380 | -0.295 | -0.241 |
| | (0.377) | (0.358) | (0.347) |
| 2002.year | -0.212 | -0.341 | -0.281 |
| | (0.373) | (0.355) | (0.350) |
| 2003.year | 0.235 | 0.126 | 0.144 |
| | (0.386) | (0.356) | (0.352) |
| 2004.year | 0.821** | 0.636* | 0.679** |
| | (0.365) | (0.341) | (0.338) |
| 2005.year | 0.751** | 0.536 | 0.528 |
| | (0.359) | (0.342) | (0.337) |
| 2006.year | 0.909** | 0.670* | 0.640* |
|  | (0.373) | (0.352) | (0.349) |
| 2007.year | 1.073*** | 0.880*** | 1.001*** |
|  | (0.346) | (0.325) | (0.322) |
| 2008.year | 1.268*** | 1.192*** | 1.337*** |
|  | (0.344) | (0.322) | (0.319) |
| 2009.year | 0.671* | 0.602* | 0.701** |
|  | (0.351) | (0.330) | (0.331) |
| 2010.year | 0.902** | 0.815** | 0.961*** |
|  | (0.356) | (0.340) | (0.338) |
| 2011.year | 1.360*** | 1.255*** | 1.359*** |
|  | (0.353) | (0.331) | (0.327) |
| 2012.year | 1.140*** | 0.964*** | 0.989*** |
|  | (0.378) | (0.365) | (0.364) |
| 2013.year | 1.268*** | 1.080*** | 1.189*** |
|  | (0.355) | (0.335) | (0.332) |
| 2014.year | 1.266*** | 1.075*** | 1.169*** |
|  | (0.336) | (0.315) | (0.316) |
| 2015.year | 1.635*** | 1.477*** | 1.572*** |
|  | (0.361) | (0.339) | (0.333) |
| 2016.year | 0.987** | 0.874** | 0.977*** |
|  | (0.388) | (0.357) | (0.355) |
| 2017.year | 1.350*** | 1.173*** | 1.297*** |
|  | (0.372) | (0.349) | (0.350) |
| 2018.year | 1.648*** | 1.387*** | 1.465*** |
|  | (0.352) | (0.331) | (0.333) |
| Constant | 8.643*** | -47.77*** | -53.36*** |
|  | (0.719) | (15.14) | (14.92) |
|  |  |  |  |
| Observations | 720 | 720 | 720 |
| Adj-R-squared | 0.459 | 0.495 | 0.510 |

**Table A2. Summary Statistics of the Information Variables for Renoir Sample**

| *Variable* | *Obs* | *Mean* | *Std. Dev.* | *Min* | *Max* |
|---|---|---|---|---|---|
| V_line | 1147 | .118 | .020 | .045 | .179 |
| V_color | 1147 | .132 | .076 | .013 | .447 |
| V_value | 1147 | .170 | .046 | .041 | .348 |
| V_shape | 1147 | .027 | .026 | .002 | .336 |
| V_space | 1147 | .135 | .017 | .096 | .189 |

| *Corr* | V_line | V_color | V_value | V_shape | V_space |
|---|---|---|---|---|---|
| V_line | 1.000 | | | | |
| V_color | -0.078 | -0.078 | | | |
| V_value | -0.414 | -0.414 | -0.414 | | |
| V_shape | 0.695 | 0.695 | 0.695 | 0.695 | |
| V_space | 0.515 | 0.515 | 0.515 | 0.515 | 0.515 |

**Table A3. Summary Statistics of other Variables for Renoir Sample**

| *Variable* | *Obs* | *Mean* | *Sd* | *Min* | *Max* |
|---|---|---|---|---|---|
| Price($) | 1147 | 675646.1 | 1491309.5 | 900 | 21000000 |
| Age | 1147 | 121.482 | 11.244 | 28 | 161 |
| Salesyear | 1147 | 2009.129 | 5.3 | 2000 | 2018 |
| Surface(1000cm$^2$) | 1147 | 1285.345 | 1512.149 | 25.806 | 22645.115 |
| Signature | 1147 | .62 | .486 | 0 | 1 |
| Dated | 1147 | .044 | .206 | 0 | 1 |
| **Material** | | | | | |
| Canvas | 1147 | 0.972 | 0.165 | 0 | 1 |
| Panel | 1147 | 0.013 | 0.114 | 0 | 1 |
| Others | 1147 | 0.148 | 0.121 | 0 | 1 |
| **City** | | | | | |
| London | 1147 | 0.347 | 0.476 | 0 | 1 |
| New York | 1147 | 0.435 | 0.495 | 0 | 1 |
| Paris | 1147 | 0.114 | 0.318 | 0 | 1 |
| Others | 1147 | 0.103 | 0.304 | 0 | 1 |
| **Salesroom** | | | | | |
| Christie's | 1147 | 0.421 | 0.494 | 0 | 1 |
| Sotheby's | 1147 | 0.399 | 0.490 | 0 | 1 |
| Others | 1147 | 0.180 | 0.384 | 0 | 1 |

Table A4. The Regression Results with all Controls of Table 5

| VARIABLES | (1) lprice | (2) lprice | (3) lprice | (4) lprice |
|---|---|---|---|---|
| *Painting Information* | | | | |
| log(V_line) | 11.23** | 8.535* | | 19.76*** |
| | (5.460) | (5.159) | | (4.423) |
| log(V_line$^2$) | -1.237** | -0.888 | | -2.270*** |
| | (0.582) | (0.546) | | (0.485) |
| log(V_color) | 0.147*** | 0.156*** | | 0.401*** |
| | (0.0365) | (0.0356) | | (0.0426) |
| log(V_value) | 0.429*** | | | 0.409*** |
| | (0.116) | | | (0.120) |
| log(V_shape) | 0.107** | | | 0.152*** |
| | (0.0495) | | | (0.0581) |
| log(V_space) | 0.837*** | | | 1.011*** |
| | (0.260) | | | (0.253) |
| *Painting Attribute* | | | | |
| Surface | 0.836*** | 0.836*** | 0.857*** | 0.148*** |
| | (0.0350) | (0.0356) | (0.0356) | (0.0133) |
| Surface$^2$ | -0.0344*** | -0.0345*** | -0.0354*** | -0.000914*** |
| | (0.00281) | (0.00268) | (0.00280) | (0.000121) |
| Age | 0.00246 | 0.00356 | 0.00395 | -0.000634 |
| | (0.00243) | (0.00248) | (0.00254) | (0.00155) |
| Signature | 0.471*** | 0.453*** | 0.481*** | 0.354*** |
| | (0.0696) | (0.0699) | (0.0722) | (0.0691) |
| Dated | 0.0242 | 0.0260 | -0.0503 | 0.841*** |
| | (0.166) | (0.168) | (0.173) | (0.0956) |
| Material | | | | |
|   canvas | 0.639** | 0.700** | 0.741** | 1.609*** |
| | (0.319) | (0.319) | (0.330) | (0.347) |
|   panel | 0.0327 | 0.0754 | 0.119 | 1.010*** |
| | (0.366) | (0.379) | (0.391) | (0.279) |
|   paper | 0.297 | 0.322 | 0.381 | 1.062*** |
| | (1.304) | (1.365) | (1.348) | (0.329) |
|   board | - | - | - | 0.699** |
| | - | - | - | (0.325) |
| *Other Control* | | | | |
| City | | | | |
|   New York | 0.421*** | 0.459*** | 0.460*** | 0.804*** |
| | (0.119) | (0.118) | (0.117) | (0.180) |
|   London | 0.571*** | 0.615*** | 0.610*** | 1.065*** |
| | (0.124) | (0.123) | (0.122) | (0.184) |
|   Paris | 0.234** | 0.277** | 0.276** | 0.285* |
| | (0.111) | (0.110) | (0.110) | (0.147) |
| Salesroom | | | | |
|   Christie's | 0.239*** | 0.196** | 0.181** | 0.436*** |
| | (0.0884) | (0.0853) | (0.0848) | (0.130) |
|   Sotheby's | 0.146 | 0.142 | 0.124 | 0.257* |
| | (0.0955) | (0.0940) | (0.0933) | (0.142) |
| Saleyear | | | | |
| 2001.year | -0.137 | -0.190 | -0.204 | -0.261 |
| | (0.158) | (0.159) | (0.153) | (0.189) |
| 2002.year | -0.0137 | -0.0446 | -0.0162 | -0.0577 |
| | (0.167) | (0.173) | (0.168) | (0.192) |
| 2003.year | 0.164 | 0.125 | 0.132 | 0.180 |
| | (0.147) | (0.149) | (0.148) | (0.189) |
| 2004.year | 0.378** | 0.351** | 0.369** | 0.877*** |
| | (0.157) | (0.159) | (0.159) | (0.190) |
| 2005.year | 0.717*** | 0.626*** | 0.618*** | 0.831*** |
| | (0.132) | (0.134) | (0.133) | (0.175) |
| 2006.year | 0.904*** | 0.843*** | 0.879*** | 0.983*** |
| | (0.125) | (0.129) | (0.129) | (0.176) |
| 2007.year | 0.919*** | 0.805*** | 0.834*** | 1.119*** |
| | (0.140) | (0.139) | (0.140) | (0.169) |
| 2008.year | 0.905*** | 0.784*** | 0.790*** | 0.999*** |
| | (0.149) | (0.151) | (0.150) | (0.174) |
| 2009.year | 0.528*** | 0.373*** | 0.401*** | 0.256 |
| | (0.131) | (0.132) | (0.130) | (0.165) |
| 2010.year | 0.772*** | 0.615*** | 0.678*** | 0.696*** |
| | (0.138) | (0.136) | (0.137) | (0.166) |
| 2011.year | 0.782*** | 0.598*** | 0.598*** | 0.998*** |
| | (0.149) | (0.152) | (0.153) | (0.180) |
| 2012.year | 0.746*** | 0.595*** | 0.631*** | 0.762*** |
| | (0.143) | (0.142) | (0.142) | (0.177) |
| 2013.year | 0.727*** | 0.592*** | 0.612*** | 0.783*** |
| | (0.145) | (0.146) | (0.143) | (0.168) |
| 2014.year | 1.040*** | 0.795*** | 0.826*** | 1.088*** |
| | (0.131) | (0.126) | (0.127) | (0.163) |
| 2015.year | 0.872*** | 0.653*** | 0.675*** | 0.939*** |
| | (0.129) | (0.126) | (0.125) | (0.178) |
| 2016.year | 0.660*** | 0.438*** | 0.447*** | 0.564*** |
| | (0.132) | (0.128) | (0.127) | (0.186) |
| 2017.year | 0.581*** | 0.406** | 0.422** | 0.702*** |
| | (0.169) | (0.167) | (0.165) | (0.183) |
| 2018.year | 0.696*** | 0.502*** | 0.541*** | 1.039*** |
| | (0.155) | (0.157) | (0.159) | (0.185) |
| Constant | -17.15 | -12.16 | 8.931*** | 10.05*** |
| | (12.81) | (12.15) | (0.471) | (0.341) |
| | | | | |
| Observations | 1,147 | 1,147 | 1,147 | 1,867 |
| Adj-R-squared | 0.602 | 0.610 | 0.621 | 0.581 |